\documentstyle[prb,aps,multicol,eqsecnum,epsf]{revtex}
\begin{document}
\draft
\tighten
\title{Parquet Graph Resummation Method for 
Vortex Liquids}
\author{
Joonhyun Yeo and M. A. Moore} 
\address{Department of Physics, University of Manchester,
Manchester, M13 9PL, United Kingdom.}
\date{\today}
\maketitle
\begin{abstract}
We present in detail a nonperturbative method for vortex
liquid systems. This method is based on the resummation of
an infinite subset of Feynman diagrams, the so-called
parquet graphs, contributing to the
four-point vertex function of the Ginzburg-Landau model for
a superconductor in a magnetic field. We derive a set of 
coupled integral equations, the parquet equations, 
governing the structure factor of the two-dimensional
vortex liquid system with and without random
impurities and the three-dimensional system in the absence of disorder.  
For the pure two-dimensional system, we simplify the parquet
equations considerably and obtain one simple equation for the structure 
factor. In two dimensions, we solve the parquet
equations numerically and find growing translational order
characterized by a length scale $R_c$ as the temperature is lowered.
The temperature dependence of $R_c$ is obtained 
in both pure and weakly disordered cases.  
The effect of disorder appears as a smooth decrease of $R_c$ as the
strength of disorder increases.
\end{abstract}
\begin{multicols}{2}

\section{introduction}
\label{sec:intro}

Since the discovery of high-$T_c$ superconductors, the nature of the 
mixed state in a type II superconductor has been a focus of 
theoretical and experimental investigation. In many high-$T_c$ materials,
thermal fluctuations are responsible for the melting
of the vortex lattice phase predicted by the mean field
theory \cite{abrikosov} into a vortex liquid phase. The presence of 
quenched random impurities in the vortex liquid or lattice phases also 
plays an important role, since it presents a possibility of 
a dissipation-free current flow due to pinning of flux lines.
New theoretical phases such as the vortex glass phase \cite{vg}
for point defects and the Bose glass phase \cite{bg} 
in the presence of extended defects have been proposed.
As shown by Larkin and Ovchinnikov (LO)
\cite{lo} the quenched point disorder destroys,
for spatial dimension $d<4$, 
the long-range crystalline order of the vortex lattice.
The system is described by 
some characteristic length 
scale $R_c$ over which a short-range translational order exists.

In a previous paper \cite{ym}, we developed a nonperturbative
scheme to calculate the structure factor of the two-dimensional (2D)
vortex liquid in the absence of random impurities. The main
ingredient of this nonperturbative method was the resummation of 
an infinite subset of Feynman diagrams contributing to the structure factor,
the so-called parquet graphs \cite{parq:gen}. This is an
analytic approach to 2D vortex liquid system which is
sophisticated enough to predict growing crystalline order in
the system as the temperature is lowered. The growth of the
translational order was investigated in connection with the
sharp peaks developing in the liquid structure factor.
Within this scheme, we found no evidence for a
finite temperature phase transition into the vortex lattice phase
and the system remains as a liquid. The length scale, $R_c$,
characterizing this growing translational order seemed to
diverge only in the zero temperature limit.

In this paper, we give detailed derivations
of the parquet graph resummation technique which were omitted
in the previous paper \cite{ym}. In addition, we 
present a new simplified version of the parquet equations governing the 
structure factor, which were given in terms of a set of 
{\em coupled} nonlinear integral equations previously. 
In the present work, we were able to obtain 
one simple equation for the structure factor, which contains all
the nonperturbative information for the 2D vortex liquid
in the absence of disorder.

In this paper, we also apply the parquet graph resummation technique to
the 2D vortex liquid in the 
presence of quenched disorder. We find that the sharp peaks which
appeared in the structure factor of the pure system become 
broadened as the strength of the disorder increases.
This is entirely consistent with one's intuition that
in the presence of disorder the length scale $R_c$ describing 
the translational order becomes smaller as compared
to the pure case. In the present work, since our
pure system is always in the liquid phase,
the effect of disorder appears as smooth deviations from the pure case.
From the nonperturbative
results, we find the temperature dependence of
the length scale $R_c$. It is, however, difficult to make any connection 
between our results and the LO type argument, since there exists no vortex
lattice phase at any finite temperature within
our nonperturbative scheme, while the LO argument always starts from
the vortex lattice with a true long-range
crystalline order.

We note that there exist 
recent theoretical studies \cite{gld} based on the elastic theory of
pinned lattices suggesting that the pinning
by quenched disorder become less effective due to the periodicity of the 
lattice so that a quasi long-range
order persist beyond the Larkin length scale $R_c$. 
In the present nonperturbative
analysis, the existence of such quasi long-range translational order 
has not been observed.

The point whether the 2D vortex liquid in the absence of disorder 
undergoes a finite temperature phase transition into a
2D vortex lattice is still controversial. Numerical 
simulations \cite{numsim} seem to suggest a first-order 
phase transition. However, as shown in Ref.~\onlinecite{noptsim},
these results depend cruicially on boundary conditions. In a
spherical geometry, the authors of Ref.~\onlinecite{noptsim}
demonstrated the absence of a finite temperature phase transition.
There also exists a recent experiment \cite{nikulov}
performed on a sample with very weak pinning, where
no sign of a phase transition is detected. The present parquet
approximation, which is an analytic theory on an infinite plane, 
seems to support the absence of a finite temperature phase transition
between 2D vortex liquid and solid.

Unlike the 2D system, it is generally believed that a 3D vortex
liquid undergoes a first-order phase transition into a vortex lattice. 
An experiment by Zeldov {\it et al.} \cite{zeldov} on BSCCO has been
accepted as a convincing evidence for the transition. However, there
exists a recent claim \cite{fetal} 
that the results of Ref.~\onlinecite{zeldov}
might be an artifact due to a particular sample geometry.  
Therefore, it would be interesting if one could develop a three-dimensional
parquet graph resummation scheme and obtain nonperturbative
information on the 3D vortex liquid system.
We find that one
can generalize the parquet equations to three dimensions
without difficulty. Unfortunately, the equations become very complicated 
and we were not able to obtain a numerical solution 
to the 3D parquet equations.

In the next section, we introduce the structure factor of a two-dimensional
disordered vortex liquid within the Ginzburg-Landau theory.
In Secs.~\ref{sec:parq1} and \ref{sec:parq2},
we present detailed derivations of the parquet equations which account
for all the parquet graphs contributing to the structure factor
for both pure and disordered cases. We also
consider zero-dimensional models to discuss the validity of the parquet
approximation in general. In the following
section, we present the main results of our calculation and discuss 
the temperature dependence of the length scale $R_c$.
In Sec.~\ref{sec:3d}, we present the generalization of the parquet 
equations to three  dimensions. Finally, we conclude with discussion
on future directions.

\section{The structure factor} 

We begin our analysis with the Ginzburg-Landau 
free energy for a superconducting film in a 
perpendicular magnetic field ${\bf B}={\bf\nabla}\times{\bf A}$ 
in the presence of quenched random impurities;
\begin{eqnarray}
F[\Psi ]&=&\int d^2{\bf r}\bigglb(
\frac{1}{2m}|(-i\hbar{\bf\nabla} -e^* {\bf A})\Psi |^2 \nonumber \\
&&~~~~~~~
+\big(\alpha+\tau({\bf r})\big) |\Psi |^2 +\frac{\beta}{2}
|\Psi |^4 \biggrb) ,
\label{fpsi}
\end{eqnarray}
where $\alpha,\beta$ and $m$ are phenomenological parameters. 
The random field $\tau ({\bf r})$ representing the quenched impurities
satisfies the probability distribution, $\overline{\tau ({\bf r})}=0$
and 
\begin{equation}
\overline{\tau ({\bf r})\tau ({\bf r}^\prime)}=\lambda
\delta^{(2)}({\bf r}-{\bf r}^\prime).
\label{tautau}
\end{equation}
In this paper we neglect the fluctuations in the vector potential
${\bf A}$ and restrict the order parameter $\Psi$ to the space spanned
by the lowest Landau level (LLL) wavefunctions.
In the symmetric gauge, where ${\bf A}=\case{B}{2}(-y,x)$, the LLL
is fully described by an arbitrary analytic function of the 
variable $z=x+iy$ multiplied by an exponential factor;
$
\Psi (x,y)=\exp (-\frac{\mu^2}{4}z^* z)\phi(z),
$
where $\mu^2\equiv e^* B/\hbar =2\pi/Q$ and $Q$ is the area of the 
unit cell of the vortex lattice. In the LLL approximation, the
GL free energy becomes
\begin{eqnarray}
F[\phi ]&=&\int dz^* dz\bigg(\big( \alpha_H +\tau (z,z^*)
\big) \exp(-\frac{\mu^2}{2}|z|^2)
|\phi (z)|^2 \nonumber \\
&&~~~~~~~~~~~~+\frac{\beta}{2}\exp (-\mu^2|z|^2)|\phi (z)|^4\bigg),
\label{fphi}
\end{eqnarray}
where $\int dz^*dz$ denotes
the integration over $x$-$y$ plane and
$\alpha_H\equiv\alpha+\hbar e^* B/2m$ vanishes at the 
mean field transition temperature.

The central quantity in this analysis is the structure factor of the
two-dimensional vortex liquid. It is proportional to the
Fourier transform of the density-density correlation
function,
$
\widetilde{\chi}({\bf k})=\int d^2{\bf R}\; e^{i{\bf k}\cdot{\bf R}}\chi
({\bf r},{\bf r}+{\bf R})$;
\[
\chi({\bf r},{\bf r}^\prime)\equiv
\overline{\langle |\Psi ({\bf r})|^2
|\Psi ({\bf r}^\prime)|^2\rangle } -\overline{\langle |\Psi ({\bf r})|^2
\rangle }\;\;
\overline{\langle |\Psi ({\bf r}^\prime)|^2\rangle },
\]
where the angular brackets denote the thermal averages.
The structure factor $\Delta ({\bf k})$ is then defined by
\begin{equation}
Q\exp
(-\frac{{\bf k}^2}{2\mu^2})\;\Delta({\bf k})\equiv
\widetilde{\chi}({\bf k})/
\bigg[ \overline{\langle | \Psi ({\bf r})|^2
\rangle}\bigg]^2 .\label{struc}
\end{equation}

A convenient way
to deal with quenched averages is to introduce $n$ replicas of 
$Z$ and calculate the correlation functions with respect to 
\begin{eqnarray}
\overline{Z^n}&=&\int\prod_a^n d\phi^*_a d\phi_a\;
\exp\bigglb[ -\int dz^* dz  \nonumber \\ 
&&\{\alpha_H e^{-\mu^2 |z|^2/2}\sum_a
|\phi_a (z)|^2 +\frac{\beta}{2} e^{-\mu^2 |z|^2}\sum_a 
|\phi_a (z)|^4 \nonumber \\
&&-\frac{\lambda}{2} e^{-\mu^2 |z|^2}\sum_{a,b}
|\phi_a (z)|^2 |\phi_b (z)|^2 \}\biggrb] \label{zn}
\end{eqnarray}
in the limit $n\rightarrow 0$. 
One can develop a standard perturbation theory for
(\ref{zn}). 
The bare propagator 
arising from the perturbation expansion of (\ref{zn})
is given by \cite{bnt}:
\begin{eqnarray}
G_0^{ab}(\zeta^*,z)&\equiv&
\langle\langle \phi^*_a (\zeta^* )\phi_b (z)\rangle\rangle_0 \nonumber \\ 
&=&\delta_{ab}
\frac{1}{\alpha_H}\frac{\mu^2}{2\pi}
\exp (\frac{\mu^2}{2}\zeta^* z),
\label{G0}
\end{eqnarray}
where the double
bracket $\langle\langle\ldots\rangle\rangle$ denotes the average 
with respect to $\overline{Z^n}$. It is important to note that, because of 
the simple quadratic term of the LLL free energy in (\ref{zn}),
the renormalized propagator $G_R^{ab}$ is obtained by simply replacing 
$\alpha_H$ by the renormalized mass $\alpha_R$. This means that
the magnetic length $\mu^{-1}$ is the only length scale 
involved in the renormalized propagator. This is one of the 
simplifying features of the LLL approximation that makes the
present nonperturbative calculation feasible.

As noticed in Ref.~\onlinecite{bnt}, the GL free energy
obtained from (\ref{zn}) is not closed under renormalization. 
The renormalization effectively generates the quartic vertices of
a general form;
\begin{eqnarray}
&&\int\prod_{i=1,2}dz^*_i dz_i e^{-\frac{\mu^2}{2}(|z_1|^2
+|z_2|^2)} \label{quartic} \\
&&\times\bigglb[
\sum_{a,b} |\phi_a(z_1)|^2 \bigg( \delta_{ab}g(|z_1 -z_2|)
-w(|z_1 - z_2| )\bigg) |\phi_b (z_2)|^2 \biggrb] ,\nonumber
\end{eqnarray}
for arbitrary functions $g$ and $w$. The bare interactions correspond to
\[
g_B (|z|)=\frac{\beta}{2}\delta (z)\delta (z^*),~~~~
w_B (|z|)=\frac{\lambda}{2}\delta (z)\delta (z^*).
\]
It is
convenient to work with the Fourier transform \cite{mn}
\[
\widetilde{g}({\bf k})=\int dz^* dz\; g (|z|)\exp\Big(\frac{i}{2}
(k^*z+kz^* )\Big)
\] 
and similarly $\widetilde{w}
({\bf k})$, where $k,k^*$ are complex momenta, $k=k_1+ik_2,
k^*=k_1-ik_2$ with ${\bf k}=(k_1,k_2)$. Thus, $\widetilde{g}_B
(| k |)=\beta /2$ and $\widetilde{w}_B
(| k |)=\lambda /2$ are constants.

In order to calculate the structure factor, (\ref{struc}), we need to 
consider the renormalized four-point correlation
function arising from Eq. (\ref{zn});
\begin{eqnarray*}
&&\langle\langle\phi^*_a (z^*_1)\phi^*_b (z^*_2)
\phi_c (z_3)\phi_d (z_4)\rangle\rangle
=G_R^{ac} (z^*_1,z_3)G_R^{bd} (z^*_2,z_4) \nonumber \\
&&+
G_R^{bc} (z^*_2,z_3)G_R^{ad} (z^*_1,z_4) 
+\langle\langle\phi^*_a (z^*_1)\phi^*_b (z^*_2)
\phi_c (z_3)\phi_d (z_4)\rangle\rangle _c,  
\end{eqnarray*}
where the last term
denotes the contribution from all connected Feynman diagrams.
To the lowest order of perturbation theory, the connected 
four-point correlation function can easily be calculated as
\begin{eqnarray}
&&\langle\langle\phi^*_a (z^*_1)\phi^*_b (z^*_2)
\phi_c (z_3)\phi_d (z_4)\rangle\rangle _c = -
\delta_{ac}\delta_{bd}\frac{2}{\alpha_R^4}(\frac{\mu^2}{2\pi})^2 
\nonumber \\
&&\times e^{\mu^2
(z^*_1 z_3+z^*_2 z_4)/2} 
\int \frac{dk^*dk}{(2\pi)^2}\Big[ \delta_{ab}
\widetilde{g}_B (|k|) -\widetilde{w}_B (|k|)\Big] \nonumber \\
&&\times e^{-|k|^2/\mu^2}e^{-i
(k^*(z_3-z_4)+k(z^*_1-z^*_2))/2}
\nonumber \\                                 
&&~~~
+(c\leftrightarrow d,~~z_3\leftrightarrow z_4 )
+O (\beta^2 , \beta\lambda , \lambda^2 ) ,\label{wa}
\end{eqnarray}
where the second term on the right hand side is the same as the first term
with $c$ and $d$, and
$z_3$ and $z_4$ interchanged.
In (\ref{wa}), we absorbed the renormalization of the propagators using the
renormalized mass $\alpha_R$. To the lowest order, one can easily
evaluate the ${\bf k}$-integrals
in (\ref{wa}), but, in general,
higher order corrections to the connected four-point function are
represented in (\ref{wa}) by the departure of the quartic vertex functions 
from constants, $\widetilde{g}_B(|k|)=\beta /2$ and
$\widetilde{w}_B(|k|)=\lambda/2$,
to general ${\bf k}$-dependent functions, $\widetilde{g}_R({\bf k})$ 
and $\widetilde{w}_R ({\bf k})$.
Therefore, the LLL approximation enables us to
concentrate on the renormalized quartic vertex function $\widetilde{g
}_R ({\bf k})$
and $\widetilde{w}_R ({\bf k})$,
which depend only on one variable instead of
three characterizing three independent channels in a usual field theory.

\begin{figure}
\narrowtext
\centerline{\epsfxsize=7cm
\epsfbox{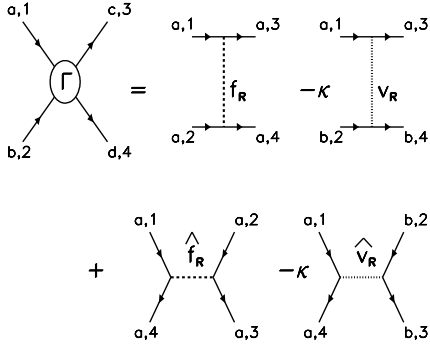}}
\vspace{5pt}
\caption{Diagrammatic representation of 
the renormalized connected four-point correlation function.
The labels, $1,\cdots , 4$, denote $z^*_1,z^*_2,z_3,z_4$, respectively,
and $a,\cdots ,d$ are replica indices.}
\label{fig:vertex}
\end{figure}

Now, one can put (\ref{wa}) for general $g_R ({\bf k})$ and
$w_R ({\bf k})$ in a more symmetric form using scaled
functions, 
\begin{eqnarray*}
&&f_R ({\bf k})\equiv \frac{2}{\beta}\exp(-{\bf k}^2/2\mu^2 )
\widetilde{g}_R ({\bf k}), \\
&&v_R ({\bf k})\equiv \frac{2}{\lambda}\exp (-{\bf k}^2/2\mu^2 )
\widetilde{w}_R ({\bf k}).
\end{eqnarray*}
The bare vertices are given by $f_B (k)=v_B (k)=\exp(-{\bf k}^2/2\mu^2)$.
Interchanging $z_3$
and $z_4$ in the second term on the right
hand side of (\ref{wa}) is equivalent to using, instead of 
$f_R ({\bf k})$ and $v_R ({\bf k})$, the transformed functions,
$\widehat{f}_R ({\bf k})$
and $\widehat{w}_R ({\bf k})$, where            
$\widehat{f} ({\bf k})$ 
is defined for an arbitrary
function $f({\bf k})$ by \cite{footnote} 
\begin{eqnarray*} 
\widehat{f} ({\bf k})\equiv\frac{2\pi}{\mu^2}\int\frac{d^2 {\bf p}}
{(2\pi)^2}\; f({\bf p})\exp\big(\frac{i}{\mu^2}(k_1 p_2 - k_2 p_1 )
\big) , \\
f({\bf k}) =\frac{2\pi}{\mu^2}\int\frac{d^2 {\bf p}}{(2\pi)^2}\;
\widehat{f}({\bf p})\exp\big(\frac{i}{\mu^2}(k_1 p_2 - k_2 p_1 )
\big) . 
\end{eqnarray*}                                  
The general four-point function is then given by (see 
Appendix)
\begin{eqnarray}
&&\langle\langle\phi^*_a (z^*_1)\phi^*_b (z^*_2)
\phi_c (z_3)\phi_d (z_4)\rangle\rangle _c = -
\frac{4}{\alpha_R^4}(\frac{\mu^2}{2\pi})^2  \label{gen:four} \\
&&\times\exp\Big(\frac{\mu^2}{2}
(z^*_1 z_3+z^*_2 z_4)\Big) 
\int \frac{dk^*dk}{(2\pi)^2}
\frac{\beta}{2}\Gamma_{ab,cd}({\bf k}) \nonumber \\
&&\times\exp\Big(-\frac{|k|^2}{2\mu^2}-\frac{i}{2}
(k^*(z_3-z_4)+k(z^*_1-z^*_2))\Big)  ,
\nonumber 
\end{eqnarray}
where 
\begin{eqnarray}
\Gamma_{ab,cd} ({\bf k})&=&
\frac{1}{2}\delta_{ac}\delta_{bd}
\Big[ \delta_{ab} f_R ({\bf k}) -\theta v_R ({\bf k})\Big] \nonumber \\
&&+ \frac{1}{2}\delta_{ad}\delta_{bc}\Big[\delta_{ab}
\widehat{f}_R ({\bf k})-\theta \widehat{v}_R ({\bf k})
\Big] ,   \label{Gff} 
\end{eqnarray}
and $\theta\equiv\lambda/\beta$ represents the strength of the 
disorder. Eq.~(\ref{Gff}) can be represented diagrammatically
as in Fig.~\ref{fig:vertex}.
                               
The structure factor $\Delta ({\bf k})$ is then obtained 
by joining two external legs of the four-point
correlation functions, (\ref{gen:four}). 
From (\ref{gen:four}) and (\ref{Gff})
and the definition (\ref{struc}) in the limit $n\rightarrow 0$, 
we obtain
\begin{equation}
\Delta({\bf k})=1- 2x \Gamma ({\bf k}), \label{dfv}
\end{equation}
where
\begin{equation}
\Gamma ({\bf k})\equiv\frac{1}{2}\bigg[ f_R ({\bf k})
+\widehat{f}_R ({\bf k})
-\theta\; \big[ v_R ({\bf k})+\widehat{v}_R ({\bf k}) \big] 
\bigg]
\end{equation}
and
$x\equiv \mu^2\beta/2\pi\alpha^2_R$ is a dimensionless parameter
which appears in the high-temeprature
perturbation expansion \cite{fhl,fn3}.

\begin{figure}
\narrowtext
\centerline{\epsfxsize=7cm\epsfbox{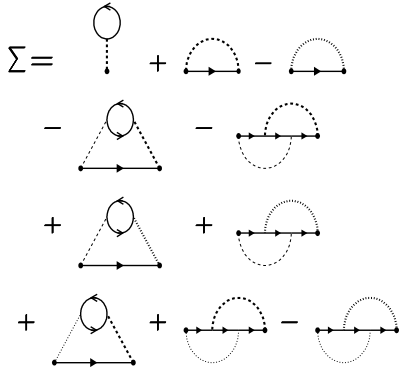}}
\vspace{5pt}
\caption{Diagrammatic representation of Dyson equation for the self-energy
$\Sigma$. The solid lines
represent the renormalized propagators.
The thick dashed and dotted lines are the renormalized 
vertex functions, $f_R$ and $v_R$, as in Fig.~1, while
the thin dashed and dotted 
lines denote the bare vertex functions, $f_B$ and $v_B$, respectively.}
\label{fig2}
\end{figure}

In the following sections, we will evaluate $f_R ({\bf k})$ and $
v_R ({\bf k})$ nonperturbatively by summing the parquet graphs. 
As mentioned earlier, we absorb any renormalization of 
the propagator into the renormalized parameter $\alpha_R$.
But knowledge of the four-point vertex function also fixes
the relationship between $\alpha_R$ and the bare parameter 
$\alpha_H$, thus completing the description of the system. 
This relation comes from the Dyson equation arising from (\ref{zn})
which is described diagrammatically in Fig.~\ref{fig2} for
the self-energy $\Sigma=G^{-1}-G^{-1}_0 =
2\pi (\alpha_R-\alpha_H)/\mu^2$:
\begin{eqnarray}
\alpha_T&=&\frac{1}{\sqrt{x}}\bigglb[ 1-x(2-\theta )+x^2 \frac{2\pi}{\mu^2}
\int\frac{d^2{\bf k}}{(2\pi)^2}\;e^{-k^2/2\mu^2}\nonumber \\
&&~~~~\times\Big[ 2(1-\theta)
f_R({\bf k})-\theta (2-\theta )v_R({\bf k})\Big]\biggrb],
\label{dyson}
\end{eqnarray}
where $\alpha_T\equiv\alpha_H\sqrt{2\pi/\beta\mu^2}$ is the dimensionless
temperature. Note that this is an {\em exact} 
relationship between the renormalized
propagator and the quartic vertex functions. We will, however,
use $f_R ({\bf k})$
and $v_R ({\bf k})$ obtained from the present nonperturbative approximation.
The Hartree approximation used
in the high-temperature perturbation expansion \cite{fhl}
corresponds to neglecting terms that depend on
the vertex functions, $f_R$ and $v_R$.

\section{Parquet Graph Resummation: Pure system ($\lambda =0$)}
\label{sec:parq1}

In order to calculate $f_R({\bf k})$ 
and $v_R ({\bf k})$, one needs to 
evaluate the Feynman diagrams contributing to the four-point correlation
function. In this and the next sections, we show in detail how to
sum an infinite subset of such diagrams, the so-called parquet graphs.

We first consider the pure system where $\lambda= 0$, then generalize the
result to the disordered case. The analysis of the pure case will mostly 
reproduce results given in Ref.\onlinecite{ym}. In the present work,
we make a further simplification of the parquet equation for $f_R ({\bf k})$
found in Ref.~\onlinecite{ym} and 
obtain a very simple equation for the structure factor $\Delta ({\bf k})$.

For both pure and disordered cases, we will present the analysis for the
zero-dimensional analogues of (\ref{fphi}), which can be integrated exactly.
For the pure case, we find it convenient to  
introduce the parquet 
resummation scheme in a simple zero-dimensional model 
and then generalize to the two-dimensional problem.
Furthermore,
since there is no apparent expansion parameter involved 
in the parquet approximation, it would be instructive to apply the parquet
resummation to the cases where exact solutions are known and to compare the
approximate result with the exact solution. 

\subsection{d=0}

\begin{figure}
\narrowtext
\centerline{\epsfxsize=7cm\epsfbox{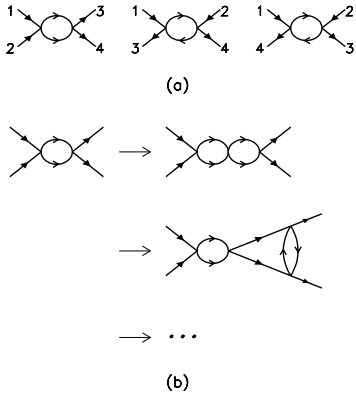}}
\vspace{1pt}
\caption{(a) Three one-loop diagrams. The symmetry factors are 2,4 and 4,
respectively. The labels $1,\cdots 4$ are drawn for the
general cases to be discussed later.
For the zero-dimensional case, there is no distinction
between the second and the third diagrams. But, it is important 
to separate this contribution into two equal parts. (b)
An example of constructing successive parquet graphs.}
\label{fig:oneloop}
\end{figure}

For $\lambda=0$ and dimension $d=0$, the partition function corresponding
to (\ref{fphi}) is a simple integral,
\begin{eqnarray*}
Z(\alpha_H,\beta)&=&\int \frac{d\psi d\psi^*}{2\pi}\;\exp (
-\alpha_H |\psi|^2-\frac{\beta}{2}|\psi|^4 ) \\
&=&\sqrt{\frac{\pi}{2\beta}}\;
{\rm erfc}(\sigma)\;\exp (\sigma^2 ), 
\end{eqnarray*}
where 
$\sigma=\alpha_H/\sqrt{2\beta}$ and
${\rm erfc}(\sigma)\equiv 1-{\rm erf}(\sigma)$ is the complementary
error function. 
In the parquet analysis, one calculates
the renormalized four-point vertex
\[
\Gamma\equiv -\frac{\alpha^4_R}{4}\langle|\psi|^4\rangle_c 
=\frac{\alpha^4_R}{2}\Big[ \frac{\partial}{\partial\beta}\ln Z
+\frac{1}{\alpha^2_R} \Big] ,
\]
where $\alpha_R^{-1}=\langle |\psi |^2 \rangle $. Although an exact solution
can readily be found, one can construct the usual Feynman graph
expansion for $\Gamma$. 
To the lowest order, we have $\Gamma=\Gamma_B =\beta/2$. To the one-loop
order, the diagrams can be represented as shown in Fig.~\ref{fig:oneloop}.
A convenient way to generate the next order diagrams is to replace 
each vertex in a one-loop diagram by the vertices obtained up to the one-loop
order. In general, one can construct higher-order
diagrams by replacing each vertex in the one-loop
diagrams by the vertices obtained up to the current order of
perturbation expansion. An example of such construction is shown in 
Fig.~\ref{fig:oneloop}. The diagrams obtained in this way are called 
{\em parquet} graphs. Note that parquet graphs can be separated into
two parts by cutting two propagator lines.

\begin{figure}
\narrowtext
\centerline{\epsfxsize=6.5cm\epsfbox{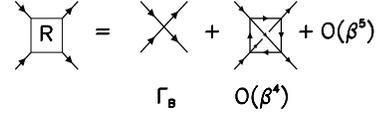}}
\vspace{1pt}
\caption{The totally irreducible vertex $R$.}
\label{fig:R}
\end{figure}

Although parquet graphs cover an enormous number of diagrams,
obviously not all diagrams can be constructed in this way. 
The non-parquet diagrams are generated in the above construction
by the so-called totally irreducible vertices whose
contribution here is denoted by $R$. The
totally irreducible vertex consists of the bare vertex and higher
order (O($\beta^4 $)) vertices (see Fig.~\ref{fig:R}).
There is no systematic ways of enumerating these higher order
diagrams contributing to $R$. The parquet approximation which we employ 
in this work corresponds to keeping only the bare vertex contribution
in $R$: $R\simeq\Gamma_B =\beta/2$ in the zero-dimensional case.
\begin{figure}
\narrowtext
\centerline{\epsfxsize=7cm\epsfbox{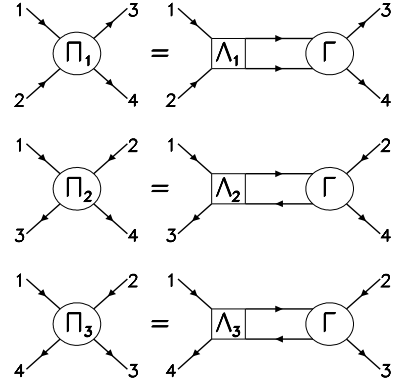}}
\vspace{1pt}
\caption{The parquet decomposition of reducible vertex $\Pi_i$.
Again, the labels $1,\cdots 4$ are drawn for the general cases.
The same diagrammatic decomposition can be used in higher dimensional models
for both pure and disordered cases.}
\label{fig:parq}
\end{figure}

If {\em both} bare vertices in a one-loop diagram are replaced by
the full renormalized vertex in the 
above construction of graphs, then the diagrams become overcounted.
There is a systematic way to eliminate this overcounting and to
generate all the Feynman diagrams and the symmetry
factors associated with each diagram, and that is to use the full vertex 
for the bare vertex on the right hand side of 
Fig.~\ref{fig:oneloop}, but the so-called {\em irreducible} vertex for the one
on the left hand side,
which is defined as follows: Let
$\Pi_i$, $i=1,2,3$ denote
the diagrams constructed out of the three one-loop diagrams
in Fig.~\ref{fig:oneloop} and $\Lambda _i$ the corresponding
irreducible vertices, we have from Fig.~\ref{fig:parq}
\begin{mathletters}
\label{g123}
\begin{eqnarray}
&&\Pi_1 =-\frac{2}{\alpha^2_R}\Lambda_1 \Gamma, \\
&&\Pi_2 =-\frac{4}{\alpha^2_R}\Lambda_2 \Gamma, \\
&&\Pi_3 =-\frac{4}{\alpha^2_R}\Lambda_3 \Gamma,
\end{eqnarray}
where we absorb any renormalization on propagator lines into
$\alpha_R$. 
Now, for the irreducible vertex $\Lambda_i$, we
must include all the diagrams that do not belong to $\Pi_i$ to avoid 
the overcounting, thus, $\Lambda_i =\Gamma -\Pi_i$.
Finally, the renormalized vertex $\Gamma$ is given by
the sum of all contributions;
\end{mathletters}
\begin{equation}
\Gamma = R + \sum_{i=1}^{3}\Pi_i .
\label{Gamma}
\end{equation}
Therefore,
\begin{equation}
\Lambda_i = R +\sum_{j\neq i}\Pi_j .
\label{Ii}
\end{equation}
Note that (\ref{g123})-(\ref{Ii}) are {\em exact}
relations for the renormalized vertex $\Gamma$. But, as mentioned before,
we use $R\simeq \beta/2$.

One can easily simplify the above equations for $R=\beta/2$ as follows: 
From (\ref{g123}), one has 
\[
\Pi_1=\frac{2\Gamma^2}{\alpha^2_R-2\Gamma},~~
\Pi_2=\Pi_3=\frac{4\Gamma^2}{\alpha^2_R-4\Gamma}.
\]
Therefore, from (\ref{Gamma}), we obtain
\begin{equation}
\gamma =\rho -\frac{\gamma^2}{1-\gamma }-2\frac{2\gamma^2}
{1-2\gamma },
\label{gamma}
\end{equation}
where $\gamma\equiv 2\Gamma/\alpha_R^2$ and $\rho\equiv\beta/\alpha_R^2$.
As a function of $\rho$, there is only one solution
$\gamma (\rho )$ that satisfies the
trivial condition: $\Gamma = \gamma = 0$ when $\beta =\rho=0$.
\begin{figure}
\narrowtext
\centerline{\epsfxsize=7cm\epsfbox{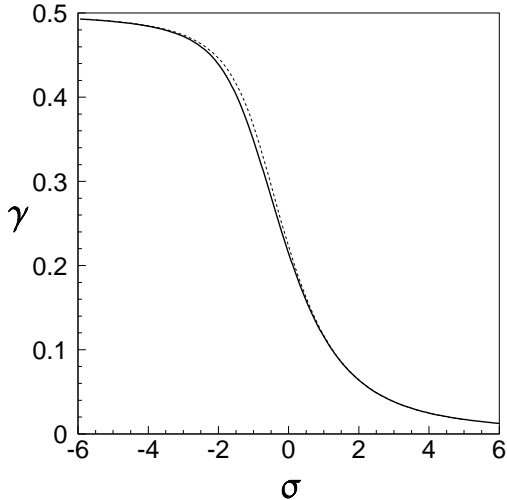}}
\vspace{5pt}
\caption{The renormalized four-point vertex function $\gamma$ for
the pure zero-dimensional model as a function of temperature $\sigma$.
The solid line is the exact solution. The dashed line
corresponds to the parquet approximation.}
\label{fig:0dp}
\end{figure}

To complete the description of the system, one needs a relation
between the bare ($\alpha_H$ or $\sigma$) and renormalized 
($\alpha_R$ or $\rho$) mass. In the parquet approximation, we 
use an analogue of Dyson equation, (\ref{dyson}), which, in this case, is
\begin{equation}
\sigma=\frac{1}{\sqrt{2\rho}}\Big[ 1- 2\rho (1-\gamma ) \Big].
\end{equation}
From this relation using $\gamma = \gamma (\rho )$ obtained from 
(\ref{gamma}), one finds $\rho$ as a function of $\sigma$, and 
consequently we have the renormalized four-point 
vertex $\gamma $ as a function of $\sigma$. In Fig.~\ref{fig:0dp},
$\gamma (\sigma)$ of the parquet approximation is compared with
the exact solution. One can see there is
excellent agreement between the two.

\subsection{d=2}

We now consider the two-dimensional problem in the absence of disorder.
The parquet equations can be constructed similarly to
the $d=0$ case. In fact, eqs.~(\ref{g123})-(\ref{Ii})
take almost the same form as before. The only differences are 
the fact that the vertices
now depend on the internal momentum ${\bf k}$ and the right hand sides of
(\ref{g123}) should be obtained from a direct evaluation of
diagrams in Fig.~\ref{fig:parq}. In Appendix, we show explicitly
how this calculation is done. The resulting equations
corresponding to (\ref{g123}) are
\begin{mathletters}
\label{P123}
\begin{eqnarray}
&&\Pi_1 ({\bf k}) = -x (\Lambda_1 \circ \Gamma ) ({\bf k}), \label{PP1}\\
&&\Pi_2 ({\bf k}) = -2x\Lambda_2 ({\bf k})\Gamma ({\bf k}), \label{PP2}\\
&&\Pi_3 ({\bf k}) = -2x( \Lambda_3 \ast \Gamma ) ({\bf k}), \label{PP3}
\end{eqnarray}
where the operations $\circ$ and $\ast$ between two arbitrary
functions $f({\bf k})$ and $g({\bf k})$ are defined by
\end{mathletters}
\begin{eqnarray*}
&& (f\circ g) ({\bf k})\equiv\frac{2\pi}{\mu^2}\int
\frac{d^2{\bf p}}{(2\pi)^2}\; f({\bf k}-{\bf p})g({\bf p})
\cos \Big(\frac{k_1 p_2 - k_2 p_1}{\mu^2}\Big) , \\
&& (f\ast g) ({\bf k})\equiv\frac{2\pi}{\mu^2}\int
\frac{d^2{\bf p}}{(2\pi)^2}\; f({\bf k}-{\bf p})g({\bf p}) .
\end{eqnarray*}
The remaining parquet equations are given by
\begin{eqnarray}
&&\Gamma ({\bf k}) = R({\bf k})+ \sum_{i=1}^3
\Pi_i ({\bf k}), \label{total} \\
&&\Lambda_i ({\bf k})=R({\bf k})
+\sum_{j\neq i}\Pi_j ({\bf k}), \label{irr}
\end{eqnarray}
where $R({\bf k})$ represents the totally irreducible part.
In the parquet approximation, $R({\bf k})$ is equal to the 
bare vertex part: $R({\bf k})\simeq
(1/2)(f_B +\widehat{f}_B)=\exp ( -{\bf k}^2/2\mu^2 )$.

We can make a
simplification on the parquet equations as 
in the zero-dimensional case. We first note the
following identities: $(f\circ g)=(\widehat{f}\circ\widehat{g})$,
$\widehat{f\circ g}=f\circ\widehat{g}=\widehat{f}\circ g$, and
$\widehat{f\ast g}({\bf k})
=\widehat{f}({\bf k})\widehat{g}({\bf k})$. Inserting $\Lambda_i ({\bf k})
=\Gamma ({\bf k}) - \Pi_i ({\bf k})$ into
(\ref{PP2}) and (\ref{PP3}) and 
using $\widehat{\Gamma}=\Gamma$ from
(\ref{Gff}), we obtain 
\begin{equation}
\Pi_2 ({\bf k})=\widehat{\Pi}_3 ({\bf k})=\frac{-2x\Gamma^2 ({\bf k})}
{1-2x\Gamma ({\bf k})}.
\label{P23}
\end{equation}
Now, inserting this into (\ref{PP1}),
\begin{eqnarray}
\Pi_1 ({\bf k})& =& -x \Big( (R + \Pi_2 +\Pi_3 )\circ \Gamma\Big)
({\bf k}) \nonumber \\
& =& -x \Big( (R +2\Pi_2 )\circ \Gamma\Big)({\bf k}). \label{P1}
\end{eqnarray}
Therefore, inserting (\ref{P23}) and (\ref{P1}) into (\ref{total}), 
we obtain an equation for $\Gamma ({\bf k})$. In terms of the structure 
factor $\Delta({\bf k})=1-2x\Gamma({\bf k})$, this equation becomes
\begin{eqnarray}
\frac{1-\Delta({\bf k})}{\Delta({\bf k})}&=& x R ({\bf k}) 
+x ( R\circ\Delta )({\bf k}) \label{main:pure} \\
&&-\bigglb(\Big[\frac
{(1-\Delta )^2}{\Delta }\Big]\circ\Delta\biggrb)({\bf k}). \nonumber
\end{eqnarray}
This is our main equation mentioned
in Sec.~\ref{sec:intro} that completely describes the structure factor of 
2D vortex liquid in the absence of disorder. Note that we have kept the
totally irreducible part $R({\bf k})$ in a general form. 
We note that, although this is an {\em exact} 
relations in a very simple form, it has little advantages for
finding numerical solutions \cite{fn:num} over the coupled parquet equations, 
(\ref{P123}), (\ref{total}) and (\ref{irr}) or another version
to be described below. 
We believe, however, that this equation might open a possibility
in the future
for a nonperturbative {\em analytic} investigation on the 2D vortex liquid.

When we consider the disordered case in the next section,
it will be {\em necessary} to decompose 
$\Gamma({\bf k})$ into $f_R ({\bf k})$ and
$\widehat{f}_R ({\bf k})$ using (\ref{Gff}) and consider the parquet 
equation in terms of $f_R ({\bf k})$ arising from the faithful representation
of Feynman diagrams as in Fig.~\ref{fig:vertex}. 
Eq.~(\ref{P23}) suggests the following decompositions:
\begin{eqnarray*}
&&\Pi_1 ({\bf k})=\frac{1}{2}\Big(\Gamma_1 ({\bf k})+\widehat{\Gamma}_1 
({\bf k})\Big) ,\\
&&\Pi_2 ({\bf k})=\widehat{\Pi}_3 ({\bf k})=
\frac{1}{2}\Big(\Gamma_2 ({\bf k})+\widehat{\Gamma}_3 ({\bf k})\Big),
\\
&&\Lambda_1 ({\bf k})=\frac{1}{2}\Big( I_1 ({\bf k})+\widehat{I}_1 ({\bf k})
\Big) , \\
&&\Lambda_2 ({\bf k})=\widehat{\Lambda}_3
({\bf k})=\frac{1}{2}\Big( I_2 ({\bf k}) +\widehat{I}_3
({\bf k})\Big) , 
\end{eqnarray*}
for some functions $I_i$ and $\Gamma_i$.
Inserting these into (\ref{P123}), we have
\begin{mathletters}
\label{original}
\begin{eqnarray}
\Gamma_1 ({\bf k}) &=&-x (I_1\circ f_R) ({\bf k}) \\
\Gamma_2 ({\bf k}) &=&-x \Big( I_2({\bf k})f_R ({\bf k})
+I_2 ({\bf k})\widehat{f}_R ({\bf k}) \nonumber \\
&&~~~~~+\widehat{I}_3 ({\bf k}) f_R ({\bf k})\Big) \\
\Gamma_3 ({\bf k}) &=&-x (I_3\ast f_R ) ({\bf k}) ,
\end{eqnarray}
and 
\begin{eqnarray}
&&f_R ({\bf k}) = f_B ({\bf k}) +\sum_{i=1}^{3}\Gamma_i ({\bf k}), \\
&&I_i ({\bf k}) = f_B ({\bf k}) +\sum_{j\neq i}\Gamma_j
({\bf k}),
\end{eqnarray}
\end{mathletters}
This version of the parquet equations
has been previously given in Ref.~\onlinecite{ym}.
 
\section{Parquet Graph Resummation: Disordered Case}
\label{sec:parq2}

For the disordered case, one has to construct the parquet equation for 
the vertex function containing $n-$replica indices.
Let us first consider the two-dimensional case directly.
It is straightforward to generalize (\ref{P123}) to the present case. 
By putting the replica indices in the diagrams in Fig.~\ref{fig:parq}, 
we have
\begin{mathletters}
\label{dis:gen}
\begin{eqnarray}
&&\Pi^{(1)}_{ab,cd}({\bf k})=-x\sum_{e,f}
\Big(\Lambda^{(1)}_{ab,ef}\circ\Gamma_{ef,cd}\Big) ({\bf k}), \\
&&\Pi^{(2)}_{ab,cd}({\bf k})=-2x\sum_{e,f}
\Lambda^{(2)}_{ae,cf}({\bf k})\Gamma_{fb,ed}({\bf k}), \\
&&\Pi^{(3)}_{ab,cd}({\bf k})=-2x\sum_{e,f}
\Big(\Lambda^{(3)}_{ae,fd}\ast\Gamma_{fb,ce}\Big) ({\bf k}).
\end{eqnarray}
The remaining equations follow from (\ref{total}) and (\ref{irr}):
\begin{eqnarray}
&&\Gamma_{ab,cd}({\bf k})=
R_{ab,cd}({\bf k})+\sum_{i=1}^{3}\Pi^{(i)}_{ab,cd}({\bf k}), \\
&&\Lambda^{(i)}_{ab,cd}({\bf k})=
R_{ab,cd}({\bf k})+\sum_{j\neq i}\Pi^{(j)}_{ab,cd}({\bf k}),
\end{eqnarray}
where
\[ 
R_{ab,cd}({\bf k})
\simeq\Gamma^{(B)}_{ab,cd}({\bf k})=\delta_{ac}\delta_{bd}(
\delta_{ab}-\theta )\exp(-{\bf k}^2/2\mu^2)
\]
in the parquet approximation.
\end{mathletters}

As one can see from (\ref{Gff}),
in order to take $n\rightarrow 0$ limit,
one needs to decompose the above equations as in the previous section
and to get equations analogous to (\ref{original}).
First, we note that 
\[
\widehat{\Lambda}^{(3)}_{ab,dc}({\bf k})=\Lambda^{(2)}_{ab,cd}
({\bf k}),~~~~\widehat{\Pi}^{(3)}_{ab,dc}({\bf k})=\Pi^{(2)}_{ab,cd}
({\bf k}).
\]
Therefore, we can write, for some functions $\Gamma_i$,
$\Xi_i$, $I_i$ and $J_i$,
\begin{eqnarray*}
\Pi^{(1)}_{ab,cd}({\bf k})
&=&\frac{1}{2}\delta_{ac}\delta_{bd}\Big[ \delta_{ab}
\Gamma_1 ({\bf k})-\theta\Xi_1 ({\bf k})\Big] \nonumber \\
&&+\frac{1}{2}\delta_{ad}\delta_{bc}\Big[ \delta_{ab}\widehat{\Gamma}_1
({\bf k})-\theta\widehat{\Xi}_1 ({\bf k})\Big], \\
\Pi^{(2)}_{ab,cd}({\bf k})&=&\widehat{\Pi}^{(3)}_{ab,dc}({\bf k})
=\frac{1}{2}\delta_{ac}\delta_{bd}\Big[ \delta_{ab}
\Gamma_2 ({\bf k})-\theta\Xi_2 ({\bf k})\Big] \nonumber \\
&&
+\frac{1}{2}\delta_{ad}\delta_{bc}\Big[ \delta_{ab}\widehat{\Gamma}_3
({\bf k})-\theta\widehat{\Xi}_3 ({\bf k})\Big],
\end{eqnarray*}
and
\begin{eqnarray*}
\Lambda^{(1)}_{ab,cd}({\bf k})
&=&\frac{1}{2}\delta_{ac}\delta_{bd}\Big[ \delta_{ab}
I_1 ({\bf k})-\theta J_1 ({\bf k})\Big] \nonumber \\
&&+\frac{1}{2}
\delta_{ad}\delta_{bc}\Big[\delta_{ab}\widehat{I}_1({\bf k})
-\theta\widehat{J}_1 ({\bf k})\Big], \\
\Lambda^{(2)}_{ab,cd}({\bf k})&=&\widehat{\Lambda}^{(3)}
_{ab,dc}({\bf k})=\frac{1}{2}\delta_{ac}\delta_{bd}\Big[\delta_{ab}
I_2 ({\bf k})-\theta J_2 ({\bf k})\Big] \nonumber \\
&&+\frac{1}{2}
\delta_{ad}\delta_{bc}\Big[\delta_{ab}\widehat{I}_3({\bf k})
-\theta\widehat{J}_3 ({\bf k})\Big].
\end{eqnarray*}

Inserting these expressions and (\ref{Gff}) 
into (\ref{dis:gen})
and eliminating one term that contains an explicit factor of $n$, we
get the following set of parquet equations 
in the presence of disorder:
\begin{eqnarray*}
\Gamma_1 ({\bf k})&=&-x(I_1\circ f_R -\theta I_1 \circ v_R
-\theta J_1\circ f_R ) ({\bf k}), \\
\Gamma_2 ({\bf k})&=&-x\Big( I_2 ({\bf k})f_R({\bf k})+I_2 ({\bf k})
\Big[
\widehat{f}_R ({\bf k})-\theta \widehat{v}_R({\bf k})\Big] \nonumber \\
&&+\Big[\widehat{I}_3({\bf k})-\theta\widehat{J}_3 
({\bf k})\Big] f_R ({\bf k})\Big) , \\
\Gamma_3 ({\bf k})&=&-x(I_3\ast f_R -\theta I_3\ast v_R -\theta
J_3 \ast f_R )({\bf k}),
\end{eqnarray*}
and
\begin{eqnarray*}
\Xi_1 ({\bf k})&=&x\theta(J_1\circ v_R )({\bf k}), \\
\Xi_2 ({\bf k})&=&-x\Big( I_2({\bf k})v_R({\bf k})+J_2({\bf k})
f_R({\bf k}) \\
&&+J_2({\bf k})\Big[ \widehat{f}_R({\bf k})-\theta 
\widehat{v}_R({\bf k})\Big] \\ 
&&+\Big[\widehat{I}_3({\bf k})-\theta\widehat{J}_3
({\bf k})\Big] v_R({\bf k})\Big), \nonumber \\
\Xi_3 ({\bf k})&=&x\theta (J_3\ast v_R)({\bf k})
\end{eqnarray*}
with
\begin{mathletters}
\label{parq:1}
\begin{eqnarray}
&&f_R ({\bf k})=f_B ({\bf k})+\sum_{i=1}^{3}\Gamma_i ({\bf k}), \\
&&v_R ({\bf k})=v_B ({\bf k})+\sum_{i=1}^{3}\Xi_i ({\bf k}),
\end{eqnarray}
and
\end{mathletters}
\begin{mathletters}
\label{parq:2}
\begin{eqnarray}
&&I_i ({\bf k})=f_B ({\bf k})+\sum_{j\neq i} \Gamma_j ({\bf k}), \\
&&J_i ({\bf k})=v_B ({\bf k})+\sum_{j\neq i} \Xi_j ({\bf k}). \label{J2}
\end{eqnarray}
\end{mathletters}

Before solving these complicated equations, we consider first the 
zero-dimensional toy model where the parquet equations reduce to
a set of algebraic equations as in (\ref{gamma}).

\subsection{d=0}

For $d=0$, the partition function is again a simple integral, 
\begin{eqnarray*}
Z(\alpha_H,\tau,\beta)&=&
\int \frac{d\psi d\psi^*}{2\pi}\; \exp\Big( -(\alpha_H +\tau ) |\psi|^2
-\frac{\beta}{2}|\psi|^4\Big)  \\
&=&\sqrt{\frac{\pi}{2\beta}}\;
{\rm erfc}\Big(\frac{\alpha_H +\tau}{
\sqrt{2\beta}}\Big)\exp\Big(\frac{(\alpha_H+\tau )^2}{2\beta}\Big).
\end{eqnarray*}
Averaging over
$\tau$ with respect to (\ref{tautau}), we have
\begin{eqnarray}
&&-\overline{\ln Z}(\sigma,\beta,\theta)=\frac{1}{2}\ln\beta
-\sigma^2-\frac{\theta}{2} \label{exact:dis} \\
&&~~~~~~~~-\frac{1}{\sqrt{\theta\pi}}\int^{\infty}
_{-\infty}d\sigma^\prime e^{-(\sigma -\sigma^\prime )^2/\theta }
\ln {\rm erfc}(\sigma^\prime ),  \nonumber
\end{eqnarray}
where
$\sigma=\alpha_H/\sqrt{2\beta}$, and $\theta=\lambda/\beta$ as
before. 

The quantities corresponding to 
$f_R ({\bf k})$ and $v_R ({\bf k})$ in the zero-dimensional case
are defined by
\begin{eqnarray*}
f_R &\equiv& -\frac{\alpha^4_R}{2\beta}
\bigglb[\overline{\langle |\psi |^4\rangle}
-2\overline{\langle |\psi |^2 \rangle^2}\biggrb] \\ 
&=&-\frac{\alpha^4_R}
{\beta}\bigg[
\frac{\partial}{\partial\beta}+\frac{\partial^2}{\partial
\alpha^2_H}\bigg]\;\overline{\ln Z} , \\
v_R &\equiv& \frac{\alpha^4_R}{\beta}
\bigglb[\overline{\langle |\psi |^2\rangle^2} -\Big[\overline{\langle
|\psi |^2\rangle}\Big]^2
\biggrb] \\
&=& -\frac{\alpha^2_R}
{\beta}-\frac{\alpha^4_R}{\beta}\bigg[ 2\frac{\partial}{\partial\beta}
+\frac{\partial^2}{\partial\alpha^2_H}\bigg]\;\overline{\ln  Z} ,
\end{eqnarray*}
where 
$\alpha^{-1}_R \equiv\overline{\langle |\psi |^2 \rangle}=-\partial
\overline{\ln Z}/\partial\alpha_H $.
These quantities can easily be evaluated as functions of $\sigma$ and
$\theta$ using (\ref{exact:dis}).

On the other hand,
the parquet equations for $f_R$ and $v_R$ take exactly the same form
as (\ref{parq:1}), (\ref{parq:2}),
if we make the following simplifications: 
the binary operations
$\circ$ and $\ast$ become multiplications between two factors; 
the transformation $\widehat{f}$ has no effect, $\widehat{f}=f$; 
$x$ is replaced by $\rho\equiv\beta/\alpha_R$; and finally $f_B=v_B=1$.
It is then straightforward to eliminate $I_i$ (or $\Gamma_i$) and
$J_i$ (or $\Xi_i$) from the
parquet equations in favor of
$f_R^\prime\equiv\rho f_R$ and $v_R^\prime
\equiv \rho v_R$. The parquet equations then reduce to the following
coupled algebraic equations.
\begin{eqnarray}
&&\frac{2f_R^\prime}{(1-f_R^\prime+v_R^\prime)(1+v_R^\prime)} \\
&&~~~~~~~+
\frac{f_R^\prime}{(1-f_R^\prime +v_R^\prime )(1-2f_R^\prime +v_R^\prime )}
-2f_R^\prime =\rho , \nonumber \\
&&\frac{2v_R^\prime}{1+v_R^\prime}+\frac{v_R^\prime}{(1-2f_R^\prime
+v_R^\prime )^2}-2v_R^\prime =\theta\rho .
\end{eqnarray}

\begin{figure}
\narrowtext
\centerline{\epsfxsize=7cm\epsfbox{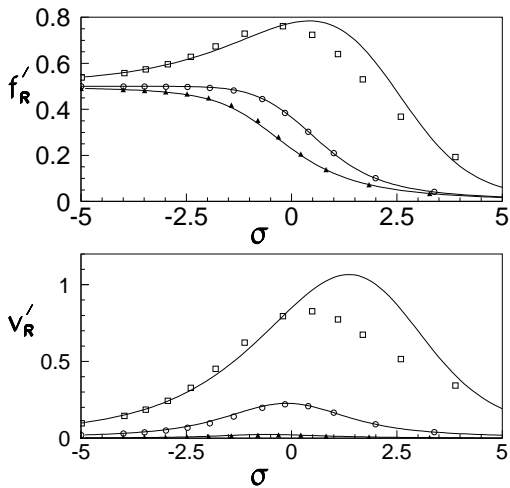}}
\vspace{10pt}
\caption{The renormalized four-point vertex functions,
$f^\prime_R$ and $v^\prime_R$ for the disordered zero-dimensional model
as functions of temperature $\sigma$. The filled triangles, circles and squares
are obtained from the parquet approximation for $\theta=0.1,1.0$ and $5.0$,
respectively. The solid lines are the corresponding exact solutions.}
\label{fig:d0d}
\end{figure}

For given $\theta$ and $\rho$, one can find the corresponding
solutions $f_R^\prime$ and 
$v_R^\prime$ to these equations. 
We can then use the Dyson equation (\ref{dyson});
\[
\sigma=\frac{1}{\sqrt{2\rho}}\Big[ 1-\rho (2-\theta ) +\rho \big[
2(f_R^\prime -v_R^\prime )-\theta (2 f_R^\prime -v_R^\prime )
\big]\Big].
\]
to find a relation between
$\rho$ and $\sigma$: $\rho=\rho (\sigma)$ for given $\theta$.
Using this, we obtain
$f_R^\prime (\sigma )$ and $v_R^\prime (\sigma )$ for given $\theta$.
These are shown in Fig.~\ref{fig:d0d} together with the exact solutions.
For $\theta\lesssim 1$, the parquet results show excellent agreement
with the exact solutions. As $\theta$ becomes larger, however, the 
discrepancies between two solutions grow. 
This analysis for the zero-dimensional model
suggests that the parquet approximation is in general 
very good when the strength of
disorder is moderate. But, the diagrams omitted
in the parquet approximation might produce quantitative errors in the strong
disorder regime. We note,
however, that physical quantities in this approximation remain
smooth as functions of $\theta$ unlike other approximation
methods \cite{bray} on this system.

\section{Results and Analysis}

For the case of the two-dimensional system, one has to solve
a set of coupled integral equations for $f_R ({\bf k})$
and $v_R ({\bf k})$, (\ref{parq:1}) and
(\ref{parq:2}), containing two parameters
$x$ and $\theta$. We consider a rotationally
symmetric case where all vertex functions depend
only on $K\equiv |{\bf k}|/\mu$. We apply a similar
numerical technique to the one used in Ref.~\onlinecite{ym}.
In practice, we find it convenient to work with $h_R (K)
\equiv f_R (K)-\theta v_R (K)$ and $v_R (K)$.
For given set of irreducible parts, $\{I_i (K)\}$ and 
$\{J_i (K)\}$, eqs.~(\ref{parq:1}) are coupled {\em linear}
integral equations for $h_R$ and $v_R$. We first solve these
equations for $h_R$ for fixed $v_R$ by numerically inverting
a matrix \cite{nr}, and then solve for $v_R$ using the solution $h_R$.
We then update the irreducible parts using (\ref{parq:2}). The solution
to the parquet equations are obtained by iterating this procedure.

A fast convergence can be obtained if we choose the initial functions,
$\{ I_i \}$, $\{J_i\}$, and $v_R$ close to the actual solutions.
At high enough temperatures, it is sufficient to start from
$I_i = J_i = v_R =\exp (-K^2/2)$. As one goes into the low temperature
regime, it is necessary to use the solution at a temperature close
to the desired temperature as initial functions. We face the same numerical 
difficulty as in Ref.~\onlinecite{ym} as the temperature is lowered, namely
one has to use a finer mesh in $K$ space as well as a larger cutoff
in order to get a low temperature solution. In this case, we have a
{\em coupled} 
set of equations, which requires an additional computing time. The
minimum temperature we used was $\alpha_T\simeq -8.5$ where the
cutoff was at $K=15$ and number of mesh points was $600$.
We also find it difficult to solve 
the parquet equations directly for arbitrarily large values 
of $\theta$. Again, one needs to start from the actual solution for 
$\theta$ close to the value for which the structure factor is 
to be calculated. In the present analysis, we were able to obtain
$\Delta(K)$ for three values of $\theta$; $\theta=0.1,0.2$ and $0.3$.

\begin{figure}
\narrowtext
\centerline{\epsfxsize=6.9cm\epsfbox{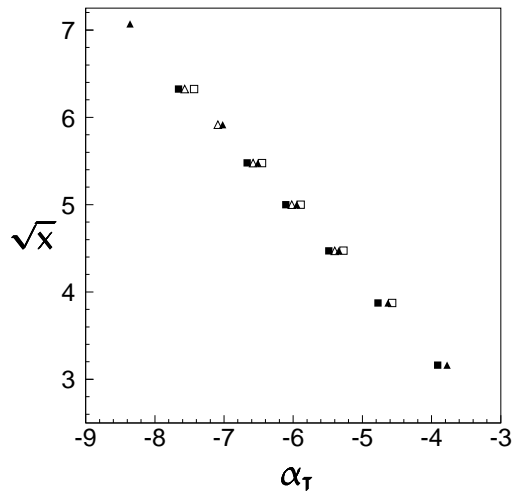}}
\vspace{5pt}
\caption{The renormalized propagator ($\mu^2/2\pi\alpha_R\sim\protect\sqrt{x}$)
as a function of
temperature $\alpha_T$ in $d=2$. The filled squares are obtained 
from the parquet approximation 
in the pure case. The open triangles, filled triangles and open squares
correspond to the parquet approximation for $\theta=0.1,0.2$ and $0.3$,
respectively.}
\label{fig:atx}
\end{figure}

In Fig.~\ref{fig:atx}, we first present the renormalized
propagator $\sqrt{x}\sim \alpha^{-1}_R$ as a function of
temperature $\alpha_T$
for three values of $\theta$.
Compared to the pure ($\theta =0$) case, $\alpha_T (x)$ as a function of $x$ 
shows very little deviations. In general, one finds that 
the same value of parameter
$x$ represents a slightly higher temperature
as $\theta$ increases.

\begin{figure}
\narrowtext
\centerline{\epsfxsize=7.5cm\epsfbox{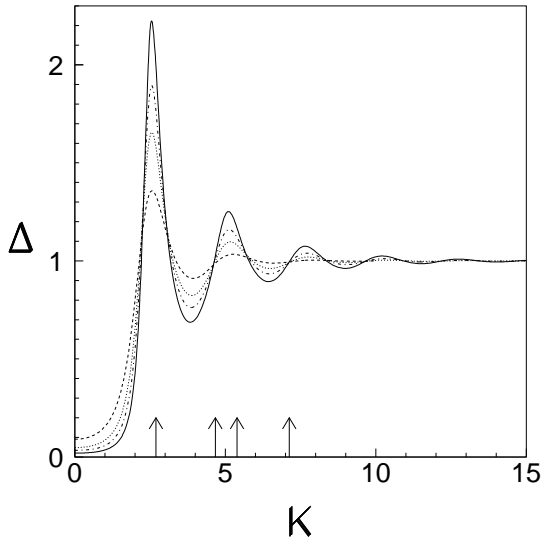}}
\vspace{10pt}
\caption{The structure factor for fixed $\theta=0.2$
and for various temperatures (dashed line: $\alpha_T=-3.78$,
dotted line: $\alpha_T=-5.34$, dot-dashed line: $\alpha_T=-6.51$
and solid line: $\alpha_T=-8.36$). The arrows indicate the positions of RLV
of the triangular lattice. Since the second and the third RLV are 
closely spaced, our solution could not resolve these peaks.}
\label{fig:struc1}
\end{figure}

\begin{figure}
\narrowtext
\centerline{\epsfxsize=7.25cm\epsfbox{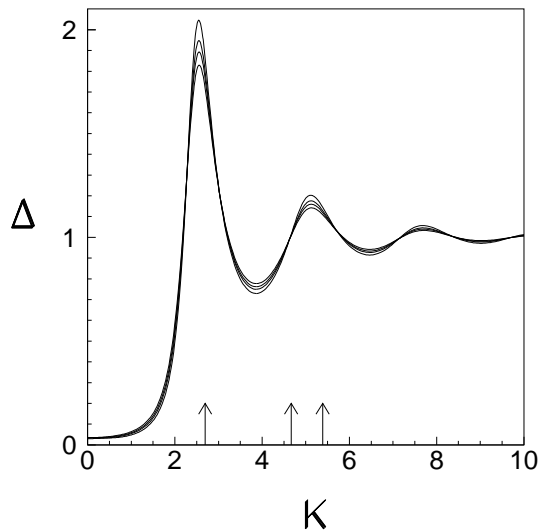}}
\vspace{10pt}
\caption{The structure factor for fixed $x=30$, (or
slightly varying $\alpha_T$; $-6.67<\alpha_T <-6.45$) 
and for various values of $\theta$: $\theta=0$ (thick line),
$0.1,0.2$ and $0.3$. The peaks are getting smaller as the strength
of disorder ($\theta$) increases.}
\label{fig:struc2}
\end{figure}

In Figs.~\ref{fig:struc1} and \ref{fig:struc2}, 
the structure factor is plotted for various
values of $\alpha_T$ and $\theta$. The 
structure factor develops a collection of peaks around the 
reciprocal lattice vectors (RLV) of the triangular lattice. 
As the temperature is lowered for fixed $\theta$, 
one can clearly see from Fig.~\ref{fig:struc1} that the first peak 
grows with decreasing width. This can be interpreted as a growing 
short-range translational order in the disordered vortex liquid. 
The length scale $R_c$ over which
this order exists can be obtained from the inverse width of the first peak,
or equivalently from the peak height \cite{ym,noptsim}. 
For fixed temperatures, we find that the
length scale decreases as the strength of disorder increases.
(See Fig.~\ref{fig:struc2}.) The result is consistent with
an intuitive picture where the disorder prevents ordering on
long length scales. In Fig.~\ref{fig:h},
we show the height of the first peak as a function of temperature for
$\theta=0.1,0.2$ and $0.3$. We find that the length scale $R_c$
grows as $|\alpha_T|$ with decreasing proportionality constant
for increasing $\theta$. 
We recall that, in the absence of disorder, the 
length scale characterizing a growing crystalline order also grows as
$|\alpha_T|$. But, from Fig.~\ref{fig:h}, we find that the rate of
this growth gets smaller as the disorder gets stronger.
In the next subsection, we show that one can derive
this behavior analytically from the parquet equations.

\subsection{Parquet Equations for $\theta\ll 1$}

For very weak disorder, $\theta\ll 1$, one can extract from
the parquet equations
the small-$\theta$ behavior of the structure factor 
$\Delta ({\bf k})$ around the pure ($\theta=0$) solution.
One can in fact derive the temperature dependence of the length 
scale $R_c$ for small $\theta$ from the parquet equations.
In order to do that, we need to find the parquet equation
analogous to (\ref{P123}), (\ref{total})
and (\ref{irr}) for $\Gamma ({\bf k})=(1/2)[h_R ({\bf k})+
\widehat{h}_R ({\bf k})]$. Out of the functions used in (\ref{parq:1})
and (\ref{parq:2}), we define 
\begin{eqnarray*}
&&\Lambda_1 ({\bf k})\equiv\frac{1}{2}\bigg[ I_1({\bf k})+\widehat{I}_1
({\bf k})-\theta\Big( J_1 ({\bf k})+\widehat{J}_1 ({\bf k})\Big)\bigg] ,\\
&&\Lambda_2 ({\bf k})\equiv\frac{1}{2}\bigg[ I_2({\bf k})+\widehat{I}_3
({\bf k})-\theta\Big( J_2 ({\bf k})+\widehat{J}_3 ({\bf k})\Big)\bigg] ,\\
&&\Lambda_3 ({\bf k})\equiv
\widehat{\Lambda}_2 ({\bf k}).
\end{eqnarray*}
and 
\begin{eqnarray*}
&&\Pi_1 ({\bf k})\equiv\frac{1}{2}\bigg[ \Gamma_1({\bf k})+
\widehat{\Gamma}_1
({\bf k})-\theta\Big( \Xi_1 ({\bf k})+\widehat{\Xi}_1 ({\bf k})\Big)\bigg], \\
&&\Pi_2 ({\bf k})\equiv\frac{1}{2}\bigg[ \Gamma_2({\bf k})+
\widehat{\Gamma}_3
({\bf k})-\theta\Big( \Xi_2 ({\bf k})+\widehat{\Xi}_3 ({\bf k})\Big)\bigg] ,\\
&&\Pi_3 ({\bf k})\equiv
\widehat{\Pi}_2 ({\bf k}).
\end{eqnarray*}
Using these functions and the original
parquet equations, (\ref{parq:1})
and (\ref{parq:2}), one can find 
a set of equations for $\Gamma ({\bf k})$ as follows:
\begin{mathletters}
\label{newparq}
\begin{eqnarray}
&&\Gamma ({\bf k})=\Gamma_B ({\bf k})+\sum_{i=1}^3\Pi_i ({\bf k}), 
\label{HR}\\
&&\Lambda_i ({\bf k})=\Gamma ({\bf k})-\Pi_i ({\bf k}),
\end{eqnarray}
\begin{eqnarray}
&&\Pi_1 ({\bf k})=-x (\Lambda\circ \Gamma)({\bf k}) ,\\
&&\Pi_2 ({\bf k})=-2x\bigg[\Lambda_2 ({\bf k})\Gamma ({\bf k})-
\frac{\theta^2}{4}J_2({\bf k})v_R ({\bf k})\bigg], \label{Pi2}\\
&&\Pi_3 ({\bf k})=\widehat{\Pi}_2({\bf k}),
\end{eqnarray}
where 
\end{mathletters}
\begin{equation}
\Gamma_B ({\bf k})=(1-\theta )\exp(-{\bf k}^2 /2\mu^2 ).
\label{GB}
\end{equation}
and we have used the above-mentioned identities
involving the binary operations, $\circ$ and $\ast$.
Note that these equations are almost the same as those 
given in the pure case, (\ref{P123}), (\ref{total}) and
(\ref{irr}). One extra term in (\ref{Pi2}) comes from
the subtraction of the diagram containg an explicit factor of $n$.
There is of course the second equation for $v_R$, which couples
nontrivially to these equations. 

Note that the structure factor $\Delta({\bf k})$ is just given by 
$\Delta=1-2x\Gamma$,
and we can simplify the above equations 
as done in the pure case to get one equation for
the structure factor.
From (\ref{Pi2}) and $J_2$ equation in (\ref{J2}), we have
\[
J_2 ({\bf k})=v_R ({\bf k})\frac{1+2x\Lambda_2 ({\bf k})}{
\Delta ({\bf k})-2\theta x v_R ({\bf k})}.
\]
Inserting this into (\ref{Pi2}), we obtain
\[
\Lambda_2 ({\bf k})=\frac{1}{2x}\bigg[-1+\frac{\Delta({\bf k})-
2\theta x v_R ({\bf k})}{[\Delta ({\bf k})-\theta x v_R ({\bf k})]^2}
\bigg].
\]

Now using (\ref{HR}) and the identities mentioned above, we find
\begin{eqnarray}
&&\frac{\Delta({\bf k})-
2\theta x v_R ({\bf k})}{[\Delta ({\bf k})-\theta x v_R ({\bf k})]^2}
=1+x \Gamma_B ({\bf k}) +x (\Gamma_B\circ\Delta)({\bf k}) \nonumber \\
&&~~~~~~-\Bigglb(\bigg[\Delta -2 + \frac{\Delta-
2\theta x v_R }{[\Delta -\theta x v_R ]^2} \bigg]\circ\Delta\Biggrb)
({\bf k}).
\label{maindisorder}
\end{eqnarray}
When $\theta=0$, this reduces to (\ref{main:pure}).
So far we have not made any assumptions on the magnitude of $\theta$.
For small $\theta$, one may consider
the lowest order perturbation in the structure
factor as compared to the pure solution. Note that, up to O($\theta$), the
equation takes the same form as in the $\theta=0$ case, 
except that $\Gamma_B({\bf k})$ in
(\ref{GB}) contains an additional factor of $(1-\theta)$.
This means that for $\theta\ll 1$,
the solution for given $\theta$ and $x$
is the same as the pure solution, denoted by $\Delta_0$, 
at $x(1-\theta)$;
\begin{equation}
\Delta({\bf k};x,\theta )\simeq \Delta_0 ({\bf k};x(1-\theta)), 
~~~\theta\ll 1.
\label{smallk}
\end{equation}
\begin{figure}
\narrowtext
\centerline{\epsfxsize=7cm\epsfbox{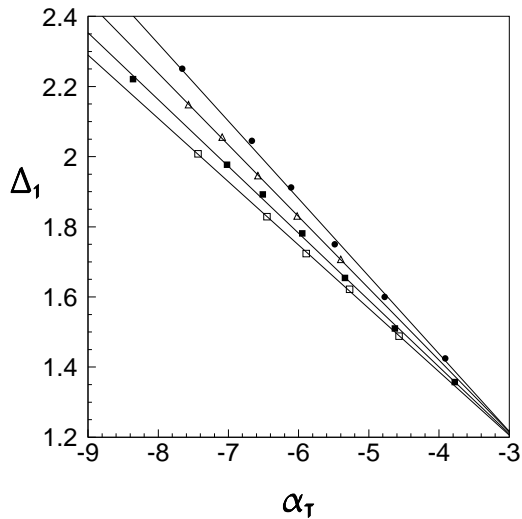}}
\vspace{10pt}
\caption{The height $\Delta_1$ of the first peak as a 
function of temperature. The filled circles, triangles, filled squares
and open squares correspond to the solutions at $\theta=0,0.1,0.2$ and
$0.3$, respectively. The solid lines are linear fits.}
\label{fig:h}
\end{figure}

For given $x$ and $\theta$, $\alpha_T$ is determined via (\ref{dyson}). 
If we denote by $\alpha^0_T (x)$ the relation between $\alpha_T$ and
$x$ for $\theta=0$, then from (\ref{dyson}) and (\ref{smallk})
we deduce that to O($\theta$)
\begin{equation}
\alpha_T (x)\simeq \alpha^0_T (x(1-\theta))-\frac{\theta}{2}
(2\sqrt{x}+\alpha^0_T (x)).
\label{atsmallk}
\end{equation}
We note that, for $\theta=0$, $\alpha^0_T (x)=(1-x\beta_A (x))/\sqrt{x}\sim
-\sqrt{x}\beta_A$ as $\alpha_T\rightarrow -\infty$, where
the generalized Abrikosov ratio changes from 2 at high temperatures
to some low temperature limit, $\beta_A$. In the pure case, we found \cite{ym}
that the length scale grows as $\sim |\alpha^0_T (x)|\sim \sqrt{x}\beta_A (x)$.
Using (\ref{smallk}) and (\ref{atsmallk}), one can derive that
\begin{equation}
R_c\sim\beta_A\sqrt{(1-\theta)x}
\sim (1-\theta (\frac{1}{2}+\frac{1}{\beta_A}))|\alpha_T| 
\label{kll1}
\end{equation}
as $\alpha_T\rightarrow -\infty$.
As mentioned before, this behavior is already captured 
in our numerical solution (see Fig.~\ref{fig:h}).

\section{Generalization to three dimensions}
\label{sec:3d}

In this section, we apply the parquet resummation method to a 
three-dimensional vortex liquid system. We shall demonstrate that,
while the generalization of the parquet equations 
to three dimensions is straightforward, 
it is virtually impossible to solve the
equations using the same numerical technique as in the 2D case.
We shall then discuss a possible analytic approach to the problem.
 
We consider only the pure case for simplicity.
The order parameter $\Psi (x,y,r_\perp)$ in 3D depends on the 
coordinate $r_\perp$ perpendicular to the $(x,y)$ plane.
In the lowest Landau level approximation, the order parameter
takes the form 
\[
\Psi (x,y,r_\perp)=\phi(z,r_\perp)\exp(-\frac{\mu^2}{4} |z|^2).
\]
The GL free energy analogous to (\ref{fphi}) is given by
\begin{eqnarray*}
F[\phi ]&=&\int dr_\perp dz^* dz\bigg(\big( 
|\partial_\perp\phi|^2+\alpha_H |\phi |^2
\big) e^{-\mu^2|z|^2/2} \nonumber \\
&&~~~~~~~~~~~~+\frac{\beta}{2}
\exp (-\mu^2|z|^2)|\phi (z,r_\perp)|^4\bigg),
\end{eqnarray*}
where $\partial_\perp = \partial/\partial r_\perp$.
The renormalized propagator can be written as
\begin{equation}
G(\zeta^*,z;q)=\frac{\mu^2}{2\pi}{\cal G}(q)\exp(\frac{\mu^2}{2}\zeta^* z),
\label{prop:3d}
\end{equation}
where $q$ is the Fourier momentum corresponding to $r_\perp$.
For the bare propagator, ${\cal G}_0(q)=(q^2+\alpha_H )^{-1}$. 
But, the renormalization drives ${\cal G}(q)$ into 
a general function. This is in contrast to the two-dimensional
case where the propagator is completely described by one parameter
$\alpha_R$.

Similarly, the renormalized four-point vertex carrys four momenta,
$q_1,q_2,q_3$, and $q_4$, in addition to ${\bf k}$ describing 
the correlation in $(x,y)$ plane. Among the four momenta
only three will be independent due to the momentum conservation
($q_1+q_2=q_3+q_4$). Using, for example, new variables,
$s\equiv q_1+q_2$, $t\equiv q_1 -q_3$ and $u\equiv q_1 -q_4$,
one can describe the renormalized four-point vertex function 
in terms of $\Gamma (s,t,u;{\bf k})$.
The parquet equations for $\Gamma$ follow from 
Fig.~\ref{fig:parq}. By evaluating
the diagrams in the Fig.~\ref{fig:parq} using the momentum conservation
on each vertex, one finds that
\end{multicols}
\widetext
\begin{eqnarray*}
&&\Pi_1 (s,t,u;{\bf k})=-\eta\int\frac{dq}{2\pi}
\Omega(q,s)\bigg[\Lambda_1 \Big(s,\frac{s+t+u}{2}-q,q-
\frac{s-t-u}{2}\Big)\circ\Gamma\Big( s, q-\frac{s-t+u}{2},
q-\frac{s+t-u}{2}\Big)\bigg]({\bf k}) \\
&&\Pi_2 (s,t,u;{\bf k})=-2\eta\int\frac{dq}{2\pi}
\Omega(q,t)\Lambda_2\Big( q+\frac{s-t+u}{2},t,\frac{s+t+u}
{2}-q;{\bf k}\Big)\Gamma\Big( q+\frac{s-t-u}{2},t,q-\frac{s+t-u}{2};
{\bf k}\Big) \\
&&\Pi_3 (s,t,u;{\bf k})=-2\eta\int\frac{dq}{2\pi}
\Omega(q,u)\bigg[\Lambda_3\Big(q+\frac{s+t-u}{2},
\frac{s+t+u}{2}-q,u\Big)\ast\Gamma\Big(q+\frac{s-t-u}{2},q-\frac{s-t+u}
{2},u\Big)\bigg]({\bf k}),
\end{eqnarray*}
\begin{multicols}{2}
\noindent where $\eta\equiv\beta\mu^2/2\pi$ and 
$\Omega(q,q^\prime)\equiv
{\cal G}(q){\cal G}(q-q^\prime )$ and $\circ$ and $\ast$ 
operate on the two-dimensional momentum ${\bf k}$ as before.
The remaining parquet equations take the same form as in the
2D case:
\begin{eqnarray*}
&&\Gamma (s,t,u;{\bf k})=R (s,t,u;{\bf k})+\sum_{i=1}^3 \Pi_i
(s,t,u;{\bf k}) \\
&&\Lambda_i (s,t,u;{\bf k})=\Gamma (s,t,u;{\bf k})-\Pi_i
(s,t,u;{\bf k}),
\end{eqnarray*}
where $R$ accounts for the totally irreducible parts, which is
set to
\[
R(s,t,u;{\bf k})\simeq\exp(-{\bf k}^2/2\mu^2)
\]
independent of $s,t,u$ in the parquet approximation.

The first difficulty one faces when one tries to solve the
above parquet equations as in the 2D case is the fact that
the propagator is given as an unknown {\em function} in this case.
In the 2D problem,  we were able to determine the unknown {\em constant}
$\alpha_R$ (or $x$) self-consistently using Dyson equation.
In this case, we have a  functional self-consistent equation for the
self-energy $\Sigma(q)={\cal G}^{-1}(q)-{\cal G}^{-1}_0(q)$ as follows:
\begin{eqnarray*}
&&\Sigma (q)= 2\eta\int\frac{dq^\prime}{2\pi}{\cal G}(q^\prime)
-2\eta^2\frac{2\pi}{\mu^2}\int\frac{d^2{\bf k}}{(2\pi)^2}
\frac{dq^\prime dq^{\prime\prime}}{(2\pi)^2}e^{-{\bf k}^2/2\mu^2}  \\
&&~\times{\cal G}(q^\prime){\cal G}(q^{\prime\prime}){\cal G}
(q+q^\prime-q^{\prime\prime})\Gamma(q+q^\prime , q-q^{\prime\prime},
q^{\prime\prime}-q^\prime;{\bf k}) 
\end{eqnarray*}

Even if one uses this self-consistent relation, the parquet equations are 
coupled integral equations for the function $\Gamma$ of four
independent variables ($s,t,u$ and $|{\bf k}|$). Storing all the data
for $\Gamma$ and for all the irreducible parts, $\Lambda_i$ in the
numerical calculation will be a formidable task. Furthermore,
the integrations in the equations become multiple sums and therefore the
previous numerical method of inverting a matrix become more cumbersome.
One may have to resort to a direct iteration, which converges slower than
the numerical inversion of matrices.

These numerical difficulties, however, should not discourage one from
attempting to get some nonperturbative information from the
parquet equations. In 2D, the parquet equations seem to be a minimal
set of equations that predicts the growing translational order
in $(x,y)$ plane, whose length scale is characterized by $R_c$.
Thus, we expect that the above 3D parquet equations contain among
other things nonperturbative information on various
length scales characterizing the low temperature regime of the system. 
In addition to $R_c$, the 3D system may have growing length scales for the 
correlation in the $r_\perp$ direction. We denote by $\xi_L$ the length scale 
arising from the propagator ${\cal G}$ and by $L_c$ the one 
associated with the 3D structure factor $S({\bf k},q)\equiv
(2\pi/\mu^2)\exp({\bf k}^2/2\mu^2)\widetilde{\chi}({\bf k},q)$ (see 
(\ref{struc})), where
\begin{eqnarray*}
&&S({\bf k},q)=\int\frac{dq^\prime}{2\pi}{\cal G}(q^\prime){\cal G}
(q+q^\prime)-2\eta\int\frac{dq^\prime dq^{\prime\prime}}{(2\pi)^2} 
{\cal G}(q^\prime){\cal G}(q^{\prime\prime}) \\
&&~~~~\times{\cal G}(q+q^\prime){\cal G}(q+q^{\prime\prime})\Gamma
(q+q^\prime+q^{\prime\prime},q,q^\prime-q^{\prime\prime};{\bf k}).
\end{eqnarray*}
If one uses in the parquet equations some kinds of
low-temperature asymptotic forms for ${\cal G}(q)$ and
$S({\bf k},q)$ which are characterized by the length scales,
$\xi_L$, $L_c$ and $R_c$, one might be able to
extract nonperturbative relations for these length scales 
and the temperature. We have tried a simple ansatz where ${\cal G}$
and $S$ are represented as delta function-like
sharp peaks around $q=0$ and ${\bf k}={\bf G}$ (RLV). The length scales are
set to be equal to the inverse width of the corresponding peaks. 
But, we found that these forms are too simplified to produce
sufficient amount of information on the temperture dependences of 
each length scale. We believe that more sophisticated asymptotic forms are 
needed for the analysis of the 3D parquet equations. This point will
be discussed again in the next section. 
   
\section{Discussion and Summary}

In this paper, we applied the parquet resummation method to
the 2D vortex liquid with and without quenched impurities and to the 3D
vortex liquid in the absence of disorder. In the 2D
system, we were able to solve the parquet equations numerically and
find the length scale $R_c$.
The temperature dependence of $R_c$ was also obtained.
In the pure 2D vortex liquid, we found in the previous paper \cite{ym}
that the asymptotic forms, $\Delta ({\bf k})\sim
2\pi\mu^2\sum_{{\bf G}\neq {\bf 0}}\delta ^{(2)} ({\bf k}-{\bf G})$,
or equivalently $f_R ({\bf k})\sim x^{-1}[1-\pi\mu^2\sum_{{\bf G}}
\delta ^{(2)}({\bf k}-{\bf G})]$ solves the parquet equations if
the inverse width of the delta-function peaks, or the length scale
$R_c$ behaves like $\sqrt{x}\sim |\alpha_T|$ as $\alpha_T
\rightarrow -\infty$. In the disordered case, one can obtain similar 
low-temperature asymptotic forms for $f_R ({\bf k})$ and $v_R ({\bf k})$
in terms of delta functions around ${\bf k}\sim{\bf G}$. One can
then obtain expressions for $R_c$ similar to (\ref{kll1}). These
simple asymptotic forms, however, have some limitations. First of all,
the amplitude of each delta-function peak cannot be determined from
the parquet equations. It is also not clear whether one can
use these representations in the strong disorder regime.
In the previous section, we argued that an ansatz like
$S({\bf k},q)\sim\delta (q)\sum_{{\bf G}\neq {\bf 0}}\delta ^{(2)}
({\bf k}-{\bf G})$ was not enough to produce any useful nonperturbative
information for the three dimensional solution. 
We believe that finding appropriate 
low temperature asymptotic forms for the vertex functions is
a first step towards a more fruitful use of the parquet equations.
To this end, a simple equation like (\ref{main:pure}) might be useful.

The effect of random impurities on the mixed state of a type-II
superconductor is usually described by the so-called Larkin-Ovchinnikov
argument \cite{lo}, which states that for a dimension $d<4$ the long-range
crystalline order of the mixed state is destroyed by a weak disorder.
For thin films ($d=2$), 
one can estimate the Larkin length scale $R_c$ over which
a short-range order persists as \cite{lo}
$R_c\sim |\alpha_T|/\lambda$. 
Although the $|\alpha_T|$-dependence of $R_c$ is the same as 
the present result, we note that there is a basic difference between
our result and the LO type argument. Since the LO argument starts from
a perfect crystalline state where $\lambda=0,R_c=\infty$, a very small 
amount of disorder makes an abrupt change as can be 
seen from $1/\sqrt{\lambda}$-
dependence of $R_c$. Within the parquet approximation, there is no 2D
crystalline phase and the pure system has a 
finite $R_c$ at any finite temperature. As we have seen in the previous
sections, the effect of small disorder is represented as a smooth
decrease of this length scale.

In this paper, we have concentrated on a rotationally symmetric liquid
state where the vertex functions depend only
on the magnitude of ${\bf k}$. But, for example,
in the presence of magnetoelastic
interactions between vortices, one must consider general 
${\bf k}$-dependent vertex functions in the parquet equations.
We recall that the input parameters of the parquet equations are the 
temperature and the bare vertex functions, $f_B$ and $v_B$. It would be 
interesting to study a situation where the bare vertex functions possess
a nontrivial ${\bf k}$-dependence. 

One could also generalize the present parquet resummation technique
to study the dynamics of type-II superconductors. We expect
that the parquet equations
for the time-dependent Ginzburg Landau theory can be obtained without
much difficulty. But, the vertex functions will depend on the
frequency as well as the usual momenta. Thus, one faces similar numerical
difficulty to the 3D case. In addition, one has to be careful in taking 
the LLL limit \cite{bm}.

\acknowledgments
We would like to thank T.\ Blum, A.\ J.\ Bray, M.\ J.\ W.\ Dodgson and
T.\ J.\ Newman for useful discussions.

\end{multicols}

\widetext
\appendix
\section*{Evaluation of diagrams}

\begin{figure}
\centerline{\epsfxsize=7cm\epsfbox{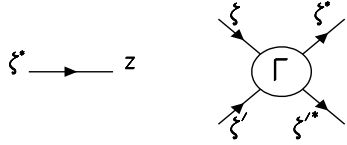}}
\vspace{10pt}
\caption{Diagrammatic representation of the propagator and the
four-point vertex.}
\label{fig:exam}
\end{figure}

Here we give some examples of evaluating the Feynman diagrams involved in
the present work. Similar analysis can be found in the Appendix
of Ref.~\onlinecite{mn:long}. We consider only the pure case 
($\lambda=0$) for simplicity. The results can easily be generalized to the
disordered case carrying the replica indices.
Feynman diagrams are constructed by connecting 
the propagator lines represented
by (\ref{G0}) with the quartic vertices, which can be obtained 
from (\ref{quartic}) as (see Fig.~\ref{fig:exam})
\[
-\frac{\beta}{2}e^{-\mu^2 (|\zeta|^2+|\zeta^\prime|^2 )/2}
\int\frac{dkdk^*}{(2\pi)^2}
\Gamma ({\bf k})e^{|k|^2/2\mu^2} 
\exp\bigg[-\frac{i}{2}\Big(k^* (\zeta -\zeta^\prime)+(\zeta^*-
\zeta^{\prime *})k\Big)
\bigg] .
\]

By joining four propagator lines (starting from
$z^*_1$ and $z^*_2$ and ending at $z_3$ and $z_4$)
with this vertex, we have
\begin{eqnarray*}
&&-4(\frac{\beta}{2})(\frac{\mu^2}{2\pi})^4\frac{1}{\alpha^4_R}
\int\frac{dkdk^*}{(2\pi )^2}\Gamma ({\bf k})e^{|k|^2/2\mu^2}
\int\prod^2_{i=1}d\zeta_i d\zeta^*_i e^{-\mu^2 (|\zeta_1|^2
+|\zeta_2|^2 )/2} \\
&&~~~\times\exp\bigglb[\frac{\mu^2}{2}\Big(z^*_1\zeta_1 +z^*_2\zeta_2
+\zeta^*_1 z_3 +\zeta^*_2 z_4\Big)-\frac{i}{2}\Big( k^* (\zeta_1
-\zeta_2 )+(\zeta^*_1 - \zeta^*_2)k \Big)\biggrb] .
\end{eqnarray*}
After performing the Gaussian integrals using
\begin{equation}
\int\prod^m_{i=1}d\zeta_i d\zeta^*_i \exp [
-\mu^2 (\zeta^*_i M_{ij} \zeta_j -a^*_i \zeta_i -\zeta ^*_i b_i )]=
(\frac{\pi}{\mu^2})^m (\det M)^{-1}\exp [ \mu^2
(a^*_i M^{-1}_{ij}b_j )] , \label{gauss}
\end{equation}
one obtains (\ref{gen:four}).

Let us evaluate the one-loop diagrams and derive (\ref{P123}). 
We consider only the first diagram of 
Fig.~\ref{fig:parq} and thus derive (\ref{PP1}). The remaining two
diagrams can be evaluated in the same manner. The first one-loop
diagram gives
\begin{eqnarray*}
&&8(\frac{\beta}{2})^2 (\frac{\mu^2}{2\pi})^6\frac{1}{\alpha^6_R}
\int\prod^2_{i=1}\frac{dk_i dk^*_i}{(2\pi)^4}\Lambda_1 ({\bf k_1})
\Gamma ({\bf k_2})e^{(|k_1|^2+|k_2|^2)/2\mu^2}\int\prod^4_{j=1}
d\zeta_j d\zeta^*_j \exp\Big(-\frac{\mu^2}{2}\sum^4_{j=1}|\zeta_j|^2
\Big) \\
&&~~~~~~~\times\exp\bigglb[ \frac{\mu^2}{2}\Big(z^*_1\zeta_1
+z^*_2\zeta_2 +\zeta^*_3 z_3 +\zeta^*_4 z_4 +\zeta^*_1\zeta_3
+\zeta^*_2\zeta_4\Big)\biggrb] \\
&&~~~~~~~\times\exp\bigglb[ -\frac{i}{2}\Big(k^*_1 (\zeta_1 -\zeta_2)
+(\zeta^*_1-\zeta^*_2)k_1+k^*_2 (\zeta_3 -\zeta_4 )+
(\zeta^*_3 -\zeta^*_4 )k_2\Big)\biggrb] .
\end{eqnarray*}
Integrating over $\zeta_i$ and $\zeta^*_i$ using (\ref{gauss}) and
changing the variables ${\bf k}_1 +{\bf k}_2 \rightarrow {\bf k}$,
and ${\bf k}_1\rightarrow {\bf p}$, we obtain
\begin{eqnarray*}
&&8(\frac{\beta}{2})^2
(\frac{\mu^2}{2\pi})^2\frac{1}{\alpha^6_R}e^{\mu^2(z^*_1 z_3 +
z^*_2 z_4)/2}\int\frac{dkdk^* dpdp^*}{(2\pi)^4}\Lambda_1 ({\bf p})
\Gamma ({\bf k}-{\bf p})e^{-|k|^2/2\mu^2} \\
&&~~~~~\exp\bigglb[-\frac{i}{2}\Big(k^* (z_3-z_4)+(z^*_1 -z^*_2)k
\Big)+\frac{1}{2\mu^2}\Big(k^* p - p^* k\Big)\biggrb] .
\end{eqnarray*}
Comparing this with the general expression (\ref{gen:four}), we find
that the one-loop contribution from this diagram to the vertex function
is 
\[
-\frac{\beta\mu^2}{2\pi\alpha^2_R}\int\frac{dpdp^*}{(2\pi)^2}
\Lambda_1 ({\bf p})\Gamma ({\bf k}-{\bf p})\exp \Big(\frac{1}{2\mu^2}
(k^* p - p^* k) \Big),
\] 
which is equal to the right hand side of (\ref{PP1}).

\begin{multicols}{2}

\end{multicols}


\begin{references}
\bibitem{footnote}
In this paper,
we assume that all the vertex functions have
a reflection symmetry; {\it e.g.} 
$f_R (-{\bf k})=f_R ({\bf k})$ so that
the exponential in the definition of $\widehat{f}_R$ 
can be replaced by a cosine. 
This is certainly true for the liquid state where the system has a
rotational symmetry and also true for a possible 
crystalline state having, for example, a triangular lattice structure.
In this work, we consider only the liquid state, but the equations 
presented here are valid for a general case with a reflection symmetry.
\bibitem{fn3}
The value of the parameter $x$ is twice the one used in
Ref.~\onlinecite{fhl} and Ref.~\onlinecite{ym}.
\bibitem{fn:num}
The numerical difficulty is mainly due to the presence of $\Delta$ in the
denominators of the equation. If one tries to solve
the equation iteratively, the procudure becomes too sensitive to the
region where $\Delta\simeq 0$.
\end{references}
\end{document}